\documentstyle[12pt]{article}
\def\ga{\gamma}
\def\de{\delta}

\def\si{\sigma}

\def\ch{\chi}
\def\ps{\psi}
\def\om{\omega}

\def\De{\Delta}

\def\Si{\Sigma}

\def\cE{{\cal E}}
\def\fr#1#2{{{#1} \over {#2}}}

\def\pt#1{\phantom{#1}}
\def\half{{\textstyle{1\over 2}}}
\def\frac#1#2{{\textstyle{{#1}\over {#2}}}}
\def\lsim{\mathrel{\rlap{\lower4pt\hbox{\hskip1pt$\sim$}}
    \raise1pt\hbox{$<$}}}
\def\gsim{\mathrel{\rlap{\lower4pt\hbox{\hskip1pt$\sim$}}
    \raise1pt\hbox{$>$}}}

\newcommand{\beq}{\begin{equation}}
\newcommand{\eeq}{\end{equation}}
\newcommand{\bea}{\begin{eqnarray}}
\newcommand{\eea}{\end{eqnarray}}
\newcommand{\rf}[1]{(\ref{#1})}
 
\begin{document}
\titlepage
 
\begin{flushright}
{IUHET 368\\}
{COLBY-97-03\\}
{August 1997\\}
\end{flushright}

\vglue 1cm
	    
\begin{center}
{{\bf CPT AND LORENTZ TESTS IN PENNING TRAPS \\ }
\vglue 1.0cm
{Robert Bluhm,$^a$ V. Alan Kosteleck\'y,$^b$ and
Neil Russell$^b$\\} 
\bigskip
{\it $^a$Physics Department, Colby College\\}
\medskip
{\it Waterville, ME 04901, U.S.A.\\}
\bigskip
{\it $^b$Physics Department, Indiana University\\}
\medskip
{\it Bloomington, IN 47405, U.S.A.\\}
}
\vglue 0.8cm
 
\end{center}
 
{\rightskip=3pc\leftskip=3pc\noindent
A theoretical analysis is performed 
of Penning-trap experiments testing CPT and Lorentz symmetry 
through measurements of anomalous magnetic moments
and charge-to-mass ratios.
Possible CPT and Lorentz violations 
arising from spontaneous symmetry breaking 
at a fundamental level
are treated in the context of a general extension 
of the SU(3)$\times$SU(2)$\times$U(1) standard model
and its restriction to quantum electrodynamics.
We describe signals that might appear in principle,
introduce suitable figures of merit, 
and estimate CPT and Lorentz bounds attainable in 
present and future Penning-trap experiments.
Experiments measuring anomaly frequencies 
are found to provide the sharpest tests of CPT symmetry.
Bounds are attainable 
of approximately $10^{-20}$ in the electron-positron case
and of $10^{-23}$ for a suggested experiment 
with protons and antiprotons.
Searches for diurnal frequency variations in these experiments
could also limit certain types of Lorentz violation
to the level of $10^{-18}$ in the electron-positron system
and others at the level of 
$10^{-21}$ in the proton-antiproton system.
In contrast,
measurements comparing cyclotron frequencies 
are sensitive within the present theoretical framework 
to different kinds of Lorentz violation
that preserve CPT.
Constraints could be obtained on one figure of merit
in the electron-positron system
at the level of $10^{-16}$,
on another in the proton-antiproton system at $10^{-24}$, 
and on a third at $10^{-25}$ using comparisons 
of $H^-$ ions with antiprotons.

}

\vskip 1 cm

\begin{center}
\it Published in Phys.\ Rev.\ D {\bf 57}, 3932-3943 (1998)
\rm 
\end{center}

%

\newpage
 
\baselineskip=20pt

{\bf\noindent I. INTRODUCTION}
\vglue 0.4cm

Invariance under the combined discrete symmetry CPT
is a fundamental symmetry of 
the SU(3)$\times$SU(2)$\times$U(1) standard model
and of quantum electrodynamics.
The CPT theorem 
\cite{cpt}
predicts that various quantities such as masses, lifetimes, 
charge-to-mass ratios, and gyromagnetic ratios
are equal for particles and antiparticles.
Typically, experimental tests of CPT are comparative measurements
of one or more of these quantities 
for a particular particle and antiparticle
\cite{pdg}.

Several high-precision tests of this type have been performed
in experiments confining single particles or antiparticles 
in a Penning trap for indefinite times.
A comparison of the electron and positron gyromagnetic ratios
can be obtained from measurements 
of their cyclotron and anomaly frequencies
\cite{vd,geo},
producing the bound
\beq
r_g \equiv |(g_- - g_+)/g_{\rm av}|
\lsim 2 \times 10^{-12}
\quad ,
\label{rg}
\eeq
where $g_-$ and $g_+$ denote the electron and positron
$g$ factors, respectively.
Similarly,
measurements of the proton and antiproton cyclotron frequencies
allow a comparison of their charge-to-mass ratios
\cite{gg1}.
The result can be presented
as the bound 
\beq
r_{q/m}^p \equiv |\left[ (q_p/m_p)
- (q_{\overline{p}}/m_{\overline{p}})
\right]/(q/m)_{\rm av}|
\lsim
1.5 \times 10^{-9}
\quad .
\label{rqmp}
\eeq
Analogous experiments performed with electrons and positrons
\cite{schwin81}
yield the bound
\beq
r_{q/m}^e \equiv |\left[ (q_{e^-}/m_{e^-})
- (q_{e^+}/m_{e^+})
\right] /(q/m)_{\rm av}|
\lsim 1.3 \times 10^{-7}
\quad .
\label{rqme}
\eeq

It has recently been shown that 
the conventional figure of merit
$r_g$ of Eq.\ \rf{rg} 
can provide a misleading measure of CPT violation
in $g - 2$ experiments
\cite{bkr}.
In the context of a general theoretical framework
that describes possible CPT- and Lorentz-violating effects
in an extension 
of the SU(3)$\times$SU(2)$\times$U(1) standard model 
and in quantum electrodynamics
\cite{ck},
the predicted value of $r_g$ is zero 
whether or not CPT is violated.
However, 
an alternative figure of merit that is 
sensitive to CPT violation does exist,
and it could be bounded to one part in $10^{20}$
with existing technology
\cite{bkr}.

In the present work,
we generalize this analysis 
to a larger class of experiments 
on charged fermions confined within a Penning trap,
including comparative measurements of 
anomaly and cyclotron frequencies
in the electron-positron, proton-antiproton,
and $H^-$-antiproton systems.
Since the dominant interactions are electromagnetic,
we consider the pure-fermion sector 
of a CPT- and Lorentz-violating extension 
of quantum electrodynamics
\cite{ck}
emerging as a limit 
of the general standard-model extension.
This broadens the scope relative to that of Ref.\ \cite{bkr},
since it also includes terms 
breaking Lorentz symmetry but preserving CPT.

Our primary goal is to determine the sensitivity
of the Penning-trap experiments 
to possible CPT- and Lorentz-violating effects
in the extension of quantum electrodynamics.
We investigate the suitability 
of the conventional figures of merit 
as measures of CPT violation.
Where necessary, 
more appropriate figures of merit and corresponding experiments
are suggested.
Estimates are also made of the magnitude of bounds 
accessible to experiments with existing technology.

Section II introduces various topics 
necessary for the analysis,
including descriptions of the 
relevant CPT- and Lorentz-violating terms,
issues concerning their perturbative treatment
in Penning-trap experiments,
and the possible signals they might engender.
Section III considers experiments with electrons and positrons
and contains three subsections:
one describing theoretical issues,
one discussing experiments on anomalous magnetic moments,
and one treating experiments on charge-to-mass ratios. 
Section IV is concerned with protons and antiprotons
and has a similar structure,
but includes a fourth subsection 
treating experiments with hydrogen ions.
We summarize in Sec.\ V.

\vglue 0.6cm
{\bf\noindent II. BASICS}
\vglue 0.4cm

{\bf\noindent A. Theoretical Framework}
\vglue 0.4cm

The framework for the extension of 
the SU(3)$\times$SU(2)$\times$U(1) standard model 
and quantum electrodynamics
originates from the idea of spontaneous CPT and Lorentz breaking
in a more fundamental model such as string theory
\cite{kp,ks}.
It lies within the context of conventional quantum field theory
and appears to preserve various desirable features
of the standard model such as gauge invariance,
power-counting renormalizability, 
and microcausality.
Possible violations of CPT and Lorentz symmetry 
are parametrized by quantities 
that can be bounded by experiments,
including interferometric tests with neutral mesons 
\cite{kp,ckpv,expt}
as well as the $g-2$ comparisons mentioned above.
There are also implications for baryogenesis
\cite{bckp}.

Within this framework,
the modified Dirac equation 
obeyed by a four-component spinor field $\ps$ 
describing a particle with charge $q$ and mass $m$
is given by
\beq
\left( i \ga^\mu D_\mu - m - a_\mu \ga^\mu
- b_\mu \ga_5 \ga^\mu - \half H_{\mu \nu} \si^{\mu \nu} 
+ i c_{\mu \nu} \ga^\mu D^\nu 
+ i d_{\mu \nu} \ga_5 \ga^\mu D^\nu \right) \ps = 0
\quad .
\label{dirac}
\eeq
Here, 
$i D_\mu \equiv i \partial_\mu - q A_\mu$,
with $A^\mu$ being the electromagnetic potential.
The quantities 
$a_\mu$, $b_\mu$, 
$H_{\mu \nu}$, $c_{\mu \nu}$, $d_{\mu \nu}$ 
are real and act as effective coupling constants,
with $H_{\mu \nu}$ antisymmetric
and $c_{\mu \nu}$, $d_{\mu \nu}$ traceless.
Some properties of these quantities 
are discussed in
Ref.\ \cite{ck}.
For our present purposes,
it suffices to note that
the transformation properties of $\ps$ 
imply the terms involving
$a_\mu$, $b_\mu$ 
break CPT
while those involving 
$H_{\mu \nu}$, $c_{\mu \nu}$, $d_{\mu \nu}$ 
preserve it,
and that Lorentz invariance is broken by all five terms.

Since no CPT or Lorentz breaking has been observed to date,
the quantities $a_\mu$, $b_\mu$, 
$H_{\mu \nu}$, $c_{\mu \nu}$, $d_{\mu \nu}$ 
must all be small.
Within the framework of
spontaneous CPT and Lorentz breaking arising
from a more fundamental model,
a natural suppression scale for these quantities 
is the ratio of a light scale
$m_l$ to a scale of order of the Planck mass $M$.
For example, 
this could range from
$m_l/M \simeq 5 \times 10^{-23}$
for $m_l \approx m_e$
to $m_l/M \simeq 3 \times 10^{-17}$
for $m_l \simeq 250$ GeV,
the latter being roughly the electroweak scale.
Since in natural units with $\hbar = c = 1$
the quantities $a_\mu$, $b_\mu$, $H_{\mu \nu}$
have dimensions of mass
while $c_{\mu \nu}$, $d_{\mu \nu}$ are dimensionless,
it is plausible that $a_\mu$, $b_\mu$, $H_{\mu \nu}$ 
might be of order $m_l m/M$,
while $c_{\mu \nu}$, $d_{\mu \nu}$ 
might be of order $m_l/M$.

\vglue 0.6cm
{\bf\noindent B. Application to the Penning Trap}
\vglue 0.4cm

The effects of the small quantities
$a_\mu$, $b_\mu$, 
$H_{\mu \nu}$, $c_{\mu \nu}$, $d_{\mu \nu}$ 
can be determined within a perturbative framework 
in relativistic quantum mechanics,
with $A_\mu$ chosen as an appropriate background potential.
The first step is therefore to extract
a suitable quantum-mechanical hamiltonian
from Eq.\ \rf{dirac}.

The appearance of time-derivative couplings in Eq.\ \rf{dirac} 
means that the standard procedure fails to produce 
a hermitian quantum-mechanical hamiltonian operator 
generating time translations on the wave function.
This technical difficulty can be overcome in several ways.
The simplest method is to perform 
a field redefinition at the lagrangian level,
chosen to eliminate the additional time derivatives.
In this case,
we find the appropriate redefinition is
\beq
\ps \equiv ( 1 - \half c_{\mu 0} \ga^0 \ga^\mu
- \half d_{\mu 0} \ga^0 \ga_5 \ga^\mu ) \ch
\quad .
\label{redef}
\eeq
Rewriting the lagrangian in terms of the new field $\ch$
cannot affect the physics.
However,
the quantum-mechanical Dirac wave function corresponding to $\ch$
does have conventional time evolution.
The physics associated with the original time-derivative couplings 
is reflected instead in additional interactions 
in the rewritten Dirac hamiltonian,
appearing as a consequence of the redefinition \rf{redef}.

We denote the Dirac wave function corresponding 
to the field $\ch$ by $\ch^q$, 
where $q \equiv e^-$ for a trapped electron 
and $q \equiv p$ for a trapped proton.
The corresponding quantum-mechanical Dirac hamiltonian
is denoted $\hat H^q$.
The rewritten Dirac equation then takes the form 
\beq
i \partial_0 \ch^q = \hat H^q \ch^q
\quad .
\label{modDirac}
\eeq
This equation
remains invariant under gauge transformations
involving $\ch^q$ and $A_\mu$.

Loop effects arising at the level of the quantum field theory
imply that the true quantum-mechanical Dirac hamiltonian
is the sum of $\hat H^q$ and other terms
that could be constructed in an effective-action approach.
In the present work,
we are interested in leading-order effects 
in the CPT- and Lorentz-violating quantities
$a_\mu$, $b_\mu$, 
$H_{\mu \nu}$, $c_{\mu \nu}$, $d_{\mu \nu}$.
We therefore work in the context of
an effective quantum-mechanical hamiltonian 
$\hat H_{\rm eff}^q$
that by definition incorporates all-orders quantum corrections
in the fine-structure constant
induced from the quantum field theory
but that keeps only first-order terms 
in CPT- and Lorentz-breaking quantities.
For perturbative calculations,
we then write
\beq
\hat H_{\rm eff}^q = \hat H_0^q + \hat H_{\rm pert}^q
\quad ,
\label{split}
\eeq
where $\hat H_0^q$ is a conventional Dirac hamiltonian 
representing a charged particle in a Penning trap 
in the absence of CPT- and Lorentz-violating perturbations 
but including quantum corrections such as an anomaly term.
The perturbative hamiltonian $\hat H_{\rm pert}^q$ 
and its analogue $\hat H_{\rm pert}^{\bar q}$
for the antiparticle
are both linear in the CPT- and Lorentz-breaking quantities
$a_\mu$, $b_\mu$, 
$H_{\mu \nu}$, $c_{\mu \nu}$, $d_{\mu \nu}$.

In a Penning trap, 
a strong magnetic field along the axis of the trap 
provides the primary radial confinement 
while axial trapping is imposed with a quadrupole electric field.
The presence of the electric field 
induces a shift in the physical cyclotron frequency
relative to its value $\om_c$ in the pure magnetic field,
but an invariance relation 
\cite{geo}
permits the value of $\om_c$ 
to be deduced directly from measurements 
of the physical cyclotron, axial, 
and magnetron frequencies in the trap.
The measurements are complicated in practice
by various experimental issues
\cite{electrn}.  
These include the disentanglement of induced couplings
between the axial and cyclotron motions,
the elimination of cyclotron-frequency shifts 
due to resonances with cavity modes inside the trap,
and the treatment of temporal drifts in the trapping fields.
Various techniques have been developed
for controlling the latter,
with accuracies of parts per billion attained
in frequency measurements 
\cite{vd,pritchard}.

For the experiments of interest here,
the dominant contributions to the energy spectrum
arise from the interaction of the particle or antiparticle
with the constant magnetic field of the trap.
Except for certain situations 
discussed in Sec. IIIA below,
the quadrupole electric and other fields 
generate smaller effects.
In a perturbative calculation,
the dominant corrections 
due to CPT- and Lorentz-violating effects
can therefore be obtained by taking $A_\mu$ 
as the potential for a constant magnetic field only. 
Since the signals of interest are energy-level shifts
rather than transition probabilities,
this means it suffices to use 
relativistic Landau-level wave functions
as the unperturbed basis set
and to calculate within first-order perturbation theory in
$\hat H_{\rm pert}^q$ or $\hat H_{\rm pert}^{\bar q}$.
However,
the unperturbed energy levels  
must be taken as the relativistic Landau levels 
shifted by an anomaly term and other quantum corrections.

As usual,
the spin-up and spin-down states form two ladders of levels.
The anomalous magnetic moment of the trapped particle
breaks the degeneracy of the excited states.
The energy-level ladder pairs for particles and antiparticles 
are similar,
except that spin labels are reversed.
Let the level number be labeled by
$n=0,1,2,3,\dots$
and the spin by 
$s = \pm 1$.
We denote the relativistic Landau-level wave functions 
for the particle and antiparticle by 
$\ch_{n,s}^q$ and $\ch_{n,s}^{\bar q}$,
respectively.
The corresponding energy levels,
including the anomaly shift 
and all conventional perturbative effects,
are denoted $E_{n,s}^q$ and $E_{n,s}^{\bar q}$.
Corrections to these energy levels 
due to CPT and Lorentz breaking are denoted by 
$\de E_{n,s}^q$ and $\de E_{n,s}^{\bar q}$
and are well approximated by 
\beq
\de E_{n,s}^{q} = \int
\ch_{n,s}^{q \dagger} \, \hat H_{\rm pert}^{q} \,
\ch_{n,s}^{q} \, d^3r
\quad , \qquad
\de E_{n,s}^{\bar q} = \int
\ch_{n,s}^{\bar q \dagger} \, \hat H_{\rm pert}^{\bar q} \,
\ch_{n,s}^{\bar q} \, d^3r
\quad .
\label{delE}
\eeq
In what follows,
the exact physical energies 
incorporating all perturbative corrections
are denoted
${\cal E}_{n,s}^q$ and ${\cal E}_{n,s}^{\bar q}$.
For calculational definiteness in the subsequent sections,
we orient the instantaneous coordinate system 
so that the magnetic field 
$\vec B = B \hat z$ lies along the positive $z$ axis,
and we choose the gauge
$A^\mu = (0,-yB,0,0)$.

To lowest order in the fine-structure constant,
we find that the perturbative hamiltonian 
$\hat H_{\rm pert}^q$ for a particle is 
\bea
\hat H_{\rm pert}^q &=& a_\mu \ga^0 \ga^\mu 
- b_\mu \ga_5 \ga^0 \ga^\mu - c_{0 0} m \ga^0
- i (c_{0 j} + c_{j 0})D^j 
+ i (c_{0 0} D_j - c_{j k} D^k) \ga^0 \ga^j
\nonumber \\
&&
- d_{j 0} m \ga_5 \ga^j + i (d_{0 j} + d_{j 0}) D^j \ga_5
+ i (d_{0 0} D_j - d_{j k} D^k) \ga^0 \ga_5 \ga^j
+ \half H_{\mu \nu} \ga^0 \si^{\mu \nu}.
\nonumber \\
\quad
\label{Hint}
\eea

For the antiparticle,
the Dirac wave function $\ch^{\bar q}$ and 
hamiltonian $\hat H^{\bar q}$ 
can be found via charge conjugation.
Experimental procedures
for replacing particles with antiparticles in Penning traps 
typically reverse the electric field 
but leave unchanged the magnetic field described by $A_\mu$.
We therefore choose the same potential $A_\mu$ in the  
Dirac hamiltonians for the particle and antiparticle.
The resulting perturbative hamiltonian 
$\hat H_{\rm pert}^{\bar q}$ for an antiparticle is 
\bea
\hat H_{\rm pert}^{\bar q} &=& - a_\mu \ga^0 \ga^\mu 
- b_\mu \ga_5 \ga^0 \ga^\mu - c_{0 0} m \ga^0
- i (c_{0 j} + c_{j 0})D^j 
+ i (c_{0 0} D_j - c_{j k} D^k) \ga^0 \ga^j
\nonumber \\
&& 
+ d_{j 0} m \ga_5 \ga^j - i (d_{0 j} + d_{j 0}) D^j \ga_5
- i (d_{0 0} D_j - d_{j k} D^k) \ga^0 \ga_5 \ga^j
- \half H_{\mu \nu} \ga^0 \si^{\mu \nu}.
\nonumber \\
\quad
\label{Hintanti}
\eea
Here,
the covariant derivative is given as
$i D_\mu = i \partial_\mu - (-q) A_\mu$,
as is appropriate for an antiparticle of charge $-q$.

In the above discussion,
the electromagnetic potential $A_\mu$ 
is treated as the usual classical background field 
solving the conventional Maxwell equations.
In principle,
effects beyond those considered here might arise 
from possible CPT- and Lorentz-breaking modifications 
of the Maxwell equations
\cite{ck}.
A plausible argument indicates that any changes 
directly involving the potential $A_\mu$ 
would be irrelevant in the situations considered here
and that the source for the extended classical theory
would still be the classical current density,
in which case a uniform magnetic field 
can be produced by conventional experimental techniques
and the results we obtain below are unaffected.
In any event,
a detailed treatment of these issues 
lies outside the scope of the present work.

\vglue 0.6cm
{\bf\noindent C. Experimental Signatures}
\vglue 0.4cm

In high-precision comparative tests 
using nonrelativistic particles or antiparticles 
confined in a Penning trap,
the relevant experimental observables are frequencies.
The effects requiring theoretical investigation
are therefore possible energy-level shifts,
which can be obtained in perturbation theory 
using Eq.\ \rf{delE}.
This subsection contains some general comments on features 
to be expected and corresponding experimental signatures.

In the present context,
the perturbative corrections to a given energy level 
could in principle depend on several variables,
including the quantum numbers of the state,
the strength of the applied field,
and its orientation.
Indeed,
all of these appear in the calculational results
presented below.

A given energy level lies in one of four stacks of levels,
according to whether the state describes a particle
or antiparticle and whether it has spin up or spin down.
Comparative tests sensitive to CPT- and Lorentz-breaking effects
could involve either states from different stacks
or states from a given stack.
For instance,
one possible effect involving different stacks
is a relative energy shift 
between particle states of one spin
and antiparticle states of the opposite spin.
The CPT theorem predicts that this difference should vanish,
assuming the trap magnetic field is the same 
for the particle and antiparticle cases.
A possible effect involving states within a given stack
is an energy shift that varies with spatial orientation.
This would conventionally be excluded by 
the rotational component of Lorentz symmetry.

The various types of CPT- and Lorentz-violating effect 
might in principle produce several kinds of observable signal
in Penning-trap experiments.
For example,
comparative measurements of anomaly frequencies could reveal 
the presence of energy-level shifts that differ
between particles and antiparticles.
Another possibility associated with level shifts
depending on spatial orientation 
is the occurrence of cyclic time variations
in either the cyclotron or anomaly frequencies.
The point is that for a given experiment
the magnetic field of the Penning trap
establishes a spatial orientation
and hence defines an instantaneous coordinate system.
This coordinate system rotates as the Earth does, 
so certain nonvanishing components of the quantities 
$a_\mu$, $b_\mu$, $H_{\mu \nu}$, $c_{\mu \nu}$, $d_{\mu \nu}$
could have time values that appear to vary diurnally
with a definite period
determined by the associated multipolarity.
Note that observing an effect would require
the absence of corresponding diurnal variations 
of the magnetic field,
which might conceivably arise 
from diurnal variations of the source 
in the effective classical Maxwell equations.
We disregard this possibility in what follows.
Note also that the magnitude of any signal would be affected
by various geometrical factors,
including the latitude 
at which the experiment is performed
and a projection of the observable 
onto the equatorial plane of the Earth.
For the order-of-magnitude estimates of bounds 
obtained in the sections that follow,
we treat these factors as being of order one. 

Since experiments measure frequencies rather than energy levels,
observable signals can only arise 
from differential energy-level shifts,
i.e.,
shifts producing changes in spacings between pairs of levels.
Furthermore,
experiments involving comparisons of frequencies between two systems
are sensitive only to double-differential level shifts,
i.e.,
level shifts that produce 
\it different \rm 
frequency shifts for each system.
The requirement of differential or double-differential level shifts 
for generation of observable signals
means that any given Penning-trap experiment
is expected to be sensitive to only a subset 
of the possible CPT- and Lorentz-breaking effects
described by Eq.\ \rf{dirac}.
This is confirmed by explicit calculation,
as is shown in the following sections.
In particular,
since the conventional figures of merit 
$r_g$, $r_{q/m}^p$, $r_{q/m}^e$
discussed in the Introduction
are defined directly 
as comparative measures of fundamental quantities,
it is unclear \it a priori \rm whether 
they are sensitive to any CPT- and Lorentz-breaking effects
and hence whether they are appropriate measures of invariance.
This question is also addressed in the following sections.

As an important example illustrating the issue of CPT sensitivity, 
consider experiments involving comparative measurements of 
cyclotron frequencies of a particle and antiparticle.
In the absence of a definite theoretical framework,
it might be expected 
\it a priori \rm
that these could reveal CPT-violating energy-level shifts.
As described above,
a CPT-breaking signal would require double-differential level shifts.
However,
there is a further constraint:
in the idealized comparative experiment
the particle and antiparticle anomaly and cyclotron frequencies 
are related not only by CPT but also by CT, 
which means that their comparison is sensitive
only to CPT-violating effects that also break CT.

In the context of the present theoretical framework,
the only terms in Eq.\ \rf{dirac} breaking both CPT and CT
are those involving the quantities $a_0$ and $\vec b$.
It has previously been shown 
\cite{ck,bkr} 
that corrections involving $a_\mu$ 
can be reinterpreted via a redefinition
of the zeros of energy and momentum,
$E \rightarrow E - a_0$ and
$\vec p \rightarrow \vec p - \vec a$,
in the dispersion relation for 
$E_{n,s}^{q} (\vec p)$.
Since all energy-level spacings 
and hence the anomaly and cyclotron frequencies remain unaffected,
these four-momentum shifts have no measurable effects 
even though the particle and antiparticle shifts 
are of opposite sign.
All observable quantities in Penning-trap experiments
are therefore independent of $a_\mu$.
To show this explicitly,
$a_\mu$ is kept in the calculations that follow.

These results imply that leading-order 
comparisons of particle and antiparticle
anomaly and cyclotron frequencies
can at most depend on $\vec b$. 
However,
the leading-order effect of a nonzero $\vec b$ 
is to shift by a constant the energy of all states with 
one spin relative to those with the other
\cite{ck,bkr}.
This means that at leading order 
a nonzero $\vec b$ is expected 
to modify anomaly-frequency comparisons
but leaves unaffected cyclotron-frequency comparisons.
In particular,
it follows that 
comparisons of particle and antiparticle cyclotron frequencies
are insensitive to all leading-order CPT-violating effects
within the present theoretical framework.

Using a related argument, 
comparative Penning-trap experiments searching for 
Lorentz-violating but CPT-preserving effects
can be shown to be sensitive 
only to effects that also preserve CT
and that couple differentially to the spin.
In the present framework,
the corresponding parameters 
are $H_{jk}$, $d_{0j}$, and $d_{j0}$.
Furthermore,
a field redefinition can be found that 
at first order in the Lorentz-breaking parameters
allows $H_{jk}$ to be absorbed into 
the antisymmetric component of $d_{j0}$
\cite{ck}.
Physical effects in the present case must therefore involve
only a particular linear combination of $H_{jk}$ 
and $d_{j0}$.
All the above results for comparative experiments 
are confirmed by the calculations that follow.

Another interesting issue is the relative sensitivity 
to possible CPT and Lorentz violation
of Penning-trap versus various other experiments.
Addressing this would require
a detailed study of the latter 
in the context of the present theoretical framework
and lies well outside the scope of the present work.
We note,
however,
that the analyses in 
Ref.\ \cite{bkr,ck,ckpv} and the following sections 
show that certain comparative Penning-trap measurements 
produce CPT bounds similar in precision
to those from experiments on neutral-meson oscillations, 
widely regarded as the best available CPT limits \cite{pdg}. 
The analysis in the present work also suggests 
that the Penning-trap sensitivity 
to possible Lorentz violation is likely to compare favorably 
with many tests of special relativity.
A few such tests,
including experiments of the Hughes-Drever type \cite{hd},
are believed under suitable circumstances
to provide exceptionally sensitive measures 
of certain kinds of Lorentz violation,
although care is required 
with interpretation of the results within specific models 
\cite{dicke}.
With some theoretical assumptions, 
these experiments might place correspondingly stringent bounds 
on parameters of interest here. 
This issue is being investigated in a separate work.

\vglue 0.6cm
{\bf\noindent III. ELECTRONS AND POSITRONS}
\vglue 0.4cm

In this section,
we consider some tests of CPT and Lorentz violation 
involving comparative experiments with
single electrons or positrons confined in a Penning trap.
The treatment is separated in three subsections,
one describing calculations of energy-level and frequency shifts,
one for experiments on anomalous magnetic moments,
and one for experiments on charge-to-mass ratios.

\vfill\eject
{\bf\noindent A. Theory}
\vglue 0.4cm

The Dirac hamiltonian $\hat H^{e^-}$
describing the electron
is identified with $\hat H^q$ of Eq.\ \rf{modDirac},
while for positrons 
$\hat H^{e^+} \equiv \hat H^{\bar q} $.
The energy levels without CPT- and Lorentz-violating perturbations
are denoted $E_{n,s}^{e^-}$ and $E_{n,s}^{e^+}$.
The corresponding electron cyclotron and anomaly frequencies
are defined as
\beq
\om_c = E_{1,-1}^{e^-} - E_{0,-1}^{e^-}
\quad ,
\quad\quad
\om_a = E_{0,+1}^{e^-} - E_{1,-1}^{e^-}
\quad .
\label{omdefs}
\eeq
By the CPT theorem,
they have the same values as those of the positron.

To distinguish 
the quantities parametrizing CPT and Lorentz breaking
for electrons and positrons 
from those for other particles introduced below,
we add superscripts:
$a_\mu^e$, $b_\mu^e$, $H_{\mu \nu}^e$, 
$c_{\mu \nu}^e$, $d_{\mu \nu}^e$.
The dominant energy-level corrections 
that are first order in these quantities 
can be calculated using Eq.\ \rf{delE}.
For electrons,
we find 
\bea
\de E_{n,\pm 1}^{e^-} &=& a_0^e
+ a_3^e \fr {p_z} {E_{n,\pm 1}^{e^-}}
\mp b_3^e 
\left( 1 - \fr {(2n + 1 \pm 1) |eB|}
{E_{n,\pm 1}^{e^-}(E_{n,\pm 1}^{e^-} + m_e)} \right)
\mp b_0^e \fr {p_z} {E_{n,\pm 1}^{e^-}}
\nonumber \\
&& \,
- c_{00}^e E_{n,\pm 1}^{e^-}
- (c_{03}^e + c_{30}^e) p_z
- (c_{11}^e + c_{22}^e) \fr {(2n + 1 \pm 1) |eB|} {2 E_{n,\pm 1}^{e^-}}
- c_{33}^e \fr {p_z^2} {E_{n,\pm 1}^{e^-}}
\nonumber \\
&& \,
\pm d_{00}^e p_z
\pm d_{30}^e m_e \left( 1 - \fr {p_z^2}
{E_{n,\pm 1}^{e^-}(E_{n,\pm 1}^{e^-} + m_e)} \right)
\pm (d_{03}^e + d_{30}^e) \fr {p_z^2} {E_{n,\pm 1}^{e^-}}
\nonumber \\
&& \,
\pm (d_{11}^e + d_{22}^e) p_z \fr {(2n + 1 \pm 1) |eB|}
{2E_{n,\pm 1}^{e^-}(E_{n,\pm 1}^{e^-} + m_e)}
\pm d_{33}^e p_z
\left( 1 - \fr {(2n + 1 \pm 1) |eB|}
{E_{n,\pm 1}^{e^-}(E_{n,\pm 1}^{e^-} + m_e)} \right)
\nonumber \\
&& \,
\pm H_{12}^e \left( 1 - \fr {p_z^2}
{E_{n,\pm 1}^{e^-}(E_{n,\pm 1}^{e^-} + m_e)} \right)
\quad .
\label{edelE}
\eea
Here,
$p_z \equiv p^3$ is the third component of the momentum.
The corresponding result for positrons,
$\de E_{n,\pm 1}^{e^+}$,
has the same structure as for the electron
but with the substitutions
$a_\mu^e \rightarrow - a_\mu^e$,
$d_{\mu \nu}^e \rightarrow - d_{\mu \nu}^e$,
$H_{\mu \nu}^e \rightarrow - H_{\mu \nu}^e$,
$E_{n,\pm 1}^{e^-} \rightarrow E_{n,\pm 1}^{e^+}$,
and $(2n + 1 \pm 1) \rightarrow (2n + 1 \mp 1)$.

In Eq.\ \rf{edelE},
corrections proportional to the magnetic field $B$
are suppressed
because the typical fields of $B \simeq 5$ T
generate only a small ratio $|eB|/m_e^2 \simeq 10^{-9}$.
Also,
axial confinement in the Penning-trap context 
is implemented by an electric field,
which means the Landau momentum $p_z$ 
appearing in Eq.\ \rf{edelE} physically corresponds
to an effective momentum for the axial motion.
The axial frequency is several orders of magnitude smaller 
than the cyclotron frequency,
so in the analysis it is tempting to neglect terms 
involving powers of the ratio $p_z / E_{n,\pm 1}^{e^-}$.
If the electric field is explicitly incorporated,
the linear terms in $p_z$ are replaced with 
expectation values involving the axial momentum.
These would vanish for stable trapping
and hence can indeed be safely ignored.
However,
in experimental situations the cooling process
can equipartition the axial and cyclotron energies,
producing large axial quantum numbers,
so that expectation values of terms quadratic in the axial momentum 
can be comparable in magnitude to the cyclotron frequency
and therefore cannot be disregarded
\it a priori. \rm
Despite this,
as is explicitly evident in the calculation that follows,
terms of this type give 
no leading-order contribution to experimental observables.

Using Eq.\ \rf{edelE},
we find that the leading-order energy corrections
are given by
\bea
\de E_{n,\pm 1}^{e^-} &\approx& a_0^e \mp b_3^e
- c_{00}^e m_e \pm d_{30}^e m_e \pm H_{12}^e
\nonumber \\ &&
- \half (c_{00}^e + c_{11}^e +c_{22}^e) (2n + 1 \pm 1) \om_c 
\nonumber \\ &&
- \left( \half c_{00}^e + c_{33}^e \mp d_{03}^e \mp d_{30}^e
\right) \fr{p_z^2}{m_e} 
\quad
\label{elecdelE}
\eea
for the electron,
and by
\bea
\de E_{n,\pm 1}^{e^+} &\approx& - a_0^e \mp b_3^e
- c_{00}^e m_e \mp d_{30}^e m_e \mp H_{12}^e
\nonumber \\ &&
- \half (c_{00}^e + c_{11}^e + c_{22}^e) (2n + 1 \mp 1) \om_c 
\nonumber \\ &&
- \left( \half c_{00}^e + c_{33}^e \pm d_{03}^e \pm d_{30}^e
\right) \fr{p_z^2}{m_e} 
\quad
\label{posdelE}
\eea
for the positron.
Keeping only resulting leading-order shifts 
in the cyclotron and anomaly frequencies 
arising from CPT and Lorentz breaking,
we find
\beq
\om_c^{e^-} \approx \om_c^{e^+} \approx
(1 - c_{00}^e - c_{11}^e - c_{22}^e) \om_c
\quad ,
\label{wcelec}
\eeq
\beq
\om_a^{e^\mp} \approx \om_a
\mp 2 b_3^e + 2 d_{30}^e m_e + 2 H_{12}^e
\quad .
\label{waelec}
\eeq
In these expressions,
$\om_c$ and $\om_a$ denote the
unperturbed frequencies
given in Eq.\ \rf{omdefs},
while $\om_c^{e^\mp}$ and $\om_a^{e^\mp}$ represent
the frequencies including the corrections.

As mentioned in Sec.\ IIC,
any cyclotron-frequency shifts 
must of necessity involve double-differential effects,
which means they depend on the quantum number $n$
and hence on the cyclotron frequency itself.
The corrections in Eq.\ \rf{wcelec}
are therefore the leading ones
in the CPT- and Lorentz-breaking quantities,
in the magnetic field,
and in the fine-structure constant.
Similarly,
Eq.\ \rf{waelec} includes all dominant terms.
For example,
the contributions to the anomaly frequencies
from Eqs.\ \rf{elecdelE} and \rf{posdelE}
that vary as $p_z^2/m_e$
are suppressed relative to the ones displayed
and hence have been omitted. 

The above derivation allows for possible relativistic effects
and quantum corrections
but treats the Penning-trap electric field only indirectly.
However,
the same result would be obtained 
from a more complete calculation.
One approach would be to treat 
the electric field and the associated axial and magnetron motions
via a Foldy-Wouthuysen diagonalization 
of the full relativistic hamiltonian.
Restricting for simplicity our attention 
to effects depending on $b_\mu^e$,
for example,
we find the contribution to the fourth-order 
Foldy-Wouthuysen hamiltonian is 
\bea
H_{b^e}^{\prime\prime\prime\prime} &=&
- \fr {b_0^e} {m_e} \vec \pi\cdot(\ga^0 \vec\Si)
- \fr {b_0^e} {2m_e^3} (\vec\pi^2 + |e|\vec B\cdot\vec\Si)
   (\vec\pi\cdot\ga^0 \vec\Si) 
\nonumber \\
&& \,
+ \vec b^e \cdot \vec \Si
+ \fr{|e|}{2m_e^3} \vec E \cdot (\vec b^e \times\vec \pi) \ga^0 
- \fr{|e|}{2m_e^2}\vec b^e \cdot(\vec B-\half i  \vec B \times\vec\Si)
\nonumber \\
&& \quad
- \fr 1 {2m_e^2} \left[(\vec b^e \cdot\vec\Si)\vec\pi^2
   - (\vec\pi\cdot\vec\Si)(\vec b^e\cdot\vec\pi)\right]
\quad .
\label{FW}
\eea
Here,
$\vec \pi = \vec p - q \vec A$ and
$\vec \Si = I \otimes \vec \si$,
where $I$ is the $2 \times 2$ unit matrix.

The hamiltonian $H_{b^e}^{\prime\prime\prime\prime}$
involves an operator momentum $\vec p$ instead 
of the constant linear momentum $p_z$.
Expectation values of the unperturbed wave functions
determine the energy shifts.
Inspection shows that neglecting the electric-field contributions 
is justified and confirms the suppression of the magnetic-field 
and other relativistic corrections compared with
the term $\vec b^e \cdot \vec \Si$,
which generates the contribution $\mp 2 b_3^e$
in Eq.\ \rf{waelec}.

The form of $H_{b^e}^{\prime\prime\prime\prime}$ 
means that terms linear in $b_0^e$ generate 
no contributions to the energy correction $\de E_{n,\pm 1}^{e^-}$,
so experiments can be sensitive at best to $(b_0^e)^2$.
In fact,
this result holds to all orders 
in the Foldy-Wouthuysen diagonalization,
as follows.
The full hamiltonian $\hat H_{\rm eff}^{e^-}$ is invariant under 
conventional parity transformations 
together with a change in sign of $b_0^e$.
The coefficient of the linear term in $b_0^e$ 
in the diagonalized hamiltonian must therefore be odd under parity.
Since parity is a symmetry of the CPT- and Lorentz-invariant
hamiltonian $\hat H_0^{e^-}$, 
the corresponding wave functions must be eigenstates of parity,
and hence the expectation values 
of terms linear in $b_0^e$ must vanish.
Note in particular that 
there are no corrections to the anomalous magnetic moment
at first order in $b_\mu^e$,
since the only term dependent on the combination
$\vec B \cdot \vec \Si$ is proportional to $b_0^e$
and produces no contribution to $\de E_{n,\pm 1}^{e^-}$.

The expressions obtained from 
a complete Foldy-Wouthuysen treatment
would depend on cyclotron, axial, 
and magnetron quantum numbers.
The present work focuses on
potentially observable shifts
in the cyclotron and anomaly frequencies,
as derived in Eqs.\ \rf{wcelec} and \rf{waelec}.
However,
we note that possible future precision experiments 
on axial or magnetron frequencies
might in principle also produce new tests 
of CPT and Lorentz symmetry.

\vglue 0.6cm
{\bf\noindent B. Anomalous Magnetic Moments}
\vglue 0.4cm

High-precision comparisons of the anomalous magnetic moments
of electrons and positrons 
\cite{vd}
currently provide the most stringent 
bounds on CPT violation in lepton systems.
These Penning-trap experiments measure 
cyclotron and anomaly frequencies 
to a precision of better than one part in $10^8$.
Combining the measurements gives the $g-2$ factors,
which are of order $10^{-3}$,
and produces the bound 
on the conventional figure of merit $r_g$
given in Eq.\ \rf{rg}.

The effects on $g-2$ measurements 
of possible CPT and Lorentz violations 
can be obtained from the results 
in the previous subsection.
Using Eqs.\ \rf{wcelec} and \rf{waelec},
we find the electron-positron differences
for the cyclotron and anomaly frequencies to be
\beq
\De \om_c^e \equiv \om_c^{e^-} - \om_c^{e^+} \approx 0
\quad , \qquad 
\De \om_a^e  \equiv \om_a^{e^-} - \om_a^{e^+} \approx - 4 b_3^e
\quad .
\label{delwce}
\eeq
The dominant signal for CPT breaking 
in Penning-trap $g-2$ experiments is therefore a difference
between the electron and positron anomaly frequencies.
No leading-order contributions appear from terms that 
preserve CPT but break Lorentz invariance.

Since the $g$ factors of the electron and positron 
are unaffected by the CPT violation to this order,
the theoretical value of $r_g$ in Eq.\ \rf{rg}
is zero whether or not CPT is broken.
Instead, 
a model-independent figure of merit providing 
a well-defined measure of CPT violation
in the weak-field, zero-momentum limit
can be introduced as
\cite{bkr}
\beq
r^e_{\om_a}
\equiv \fr{|{\cal E}_{n,s}^{e^-} - {\cal E}_{n,-s}^{e^+}|}
{{\cal E}_{n,s}^{e^-} }
\quad .
\label{re}
\eeq
Within the present framework for CPT violation,
it can be shown that 
\beq
r^e_{\om_a} \approx 
| \De \om_a^e | / 2 m_e  \approx |2 b_3^e | / m_e
\quad .
\label{reb}
\eeq
Note that since the frequency difference $\De \om_a^e $ 
depends only on the projection of $\vec b^e$ along $\hat B$
while the direction of $\hat B$ can be changed,
bounds on different spatial components of $\vec b^e$ are
possible in principle.
With the cyclotron frequency as a magnetometer,
experiments using existing techniques
could place an estimated bound on this figure of merit 
\cite{bkr}:
\beq
r^e_{\om_a} \lsim 10^{-20}
\quad .
\label{relim}
\eeq

As mentioned in Sec.\ IIC,
there exists another class of possible experimental signal,
involving a diurnal variation of anomaly-frequency measurements.
In particular,
the energy corrections $\de E_{n,\pm 1}^{e^-}$ and
$\de E_{n,\pm 1}^{e^+}$ 
could change as the Earth rotates,
producing variations in $\om_c^{e^\mp}$ and $\om_a^{e^\mp}$ 
in Eqs.\ \rf{wcelec} and \rf{waelec}.
However,
$g-2$ experiments typically determine the ratio
$2 \om_a^{e^\mp}/\om_c^{e^\mp}$
rather than obtaining absolute measurements of $\om_a^{e^\mp}$.
This avoids problems with drifting magnetic fields.
Using the cyclotron frequency 
for controlling and monitoring such drifts
in a search for diurnal variations
is problematic in principle 
since it too could contain signal time variations,
as might other possible monitoring devices.

Nonetheless,
even under circumstances where
sizable field drifts cannot be excluded,
a relatively stringent bound on Lorentz violation can be obtained.
Consider the average 
$(\om_a^{e^-} + \om_a^{e^+})/2$
of the electron and positron anomaly frequencies.
Using Eq.\ \rf{waelec} with equal magnetic fields,
we find 
\beq
\half (\om_a^{e^-} + \om_a^{e^+}) \approx
\om_a + 2 d_{30}^e m_e + 2 H_{12}^e
\quad .
\label{addwa}
\eeq
Suppose field-drift effects,
including systematic effects such as diurnal temperature changes,
cannot be excluded,
and assume no significant Lorentz violation is detected.
Then,
as electrons and positrons are alternately loaded 
in the Penning trap during the course of the experiment,
we conservatively estimate that
the time variation of the measured value of the 
anomaly-frequency average 
would be confined at least to within a 1 kHz band 
centered on the mean value.
This corresponds to a maximal field drift 
limited to 5 parts in $10^6$
for the typical superconducting solenoids used.

As before,
a suitable model-independent figure of merit
can be introduced theoretically in terms of 
differences between exact energy levels.
Define
\beq
\De^e_{\om_a} \equiv 
\fr {|{\cE}_{0,+1}^{e^-} - {\cE}_{1,-1}^{e^-}|}
{2{\cE}_{0,-1}^{e^-}} 
+
\fr {|{\cE}_{0,-1}^{e^+} - {\cE}_{1,+1}^{e^+}|}
{2{\cE}_{0,+1}^{e^+}} 
\quad .
\label{Delediurnal}
\eeq
If diurnal variations arise due to Lorentz-violating effects,
then $\De^e_{\om_a}$ would display a periodic time dependence.
The appropriate figure of merit would be 
the (dimensionless) amplitude of this oscillation,
which we denote
$r^e_{\om_a, \rm diurnal}$.
In the context of the present framework,
we find using Eqs.\ \rf{addwa} and \rf{Delediurnal}
that this figure of merit
depends on a combination of Lorentz-violating quantities,
\beq
r^e_{\om_a, \rm diurnal} \approx
|d_{30}^e m_e + H_{12}^e |/m_e 
\quad ,
\label{dHliman}
\eeq
expressed in the comoving laboratory frame on the Earth.
The restriction to a 1 kHz band mentioned above 
then yields an estimated bound of 
\beq
r^e_{\om_a, \rm diurnal} \lsim 10^{-18}
\quad .
\label{dHlim}
\eeq
With magnetic fields stable to one part in $10^9$,
a thousandfold improvement in this bound would be plausible.

\vglue 0.6cm
{\bf\noindent C. Charge-to-Mass Ratios}
\vglue 0.4cm

Experiments measuring cyclotron frequencies 
also provide high-precision comparisons 
of isolated electrons and positrons confined in a Penning trap.
These measurements are conventionally interpreted 
as determining charge-to-mass ratios.
The associated conventional figure of merit,
given in Eq.\ \rf{rqme},
is related to experimentally
measured quantities by $r_{q/m}^e = |\De \om_c^e/\om_c^{e^-}|$,
where $\De \om_c^e$ is the electron-positron 
cyclotron-frequency difference.

The present theoretical framework 
for treating CPT and Lorentz violation 
can be used to examine possible effects 
on the electron and positron cyclotron frequencies.
These acquire corrections given in Eq.\ \rf{wcelec}.
An immediate result is that to leading order 
the frequencies $\om_c^{e^\mp}$ 
are independent of CPT-violating quantities.
Since the electron and positron cyclotron frequencies 
can remain unchanged even in the presence of CPT violation,
it would be misleading to regard comparisons 
of these frequencies as appropriate measures of CPT breaking.
In particular,
this applies to the figure of merit 
$r_{q/m}^e$ in Eq.\ \rf{rqme},
which is controlled by the frequency difference $\De \om_c^e$.

The leading-order cyclotron-frequency shifts 
in Eq.\ \rf{wcelec} do display dependence 
on the Lorentz-breaking but CPT-preserving quantity 
$c_{\mu \nu}^e$.
However,
the instantaneous equality 
of the electron and positron cyclotron frequencies 
means that it would also be misleading to regard their difference 
as an appropriate signal for Lorentz violation.

Another possibility is to search for 
diurnal variations in either $\om_c^{e^-}$ or $\om_c^{e^+}$,
which might arise from the dependence of these frequencies 
on the combination of spatial components 
$|c_{11}^e + c_{22}^e|$ of $c_{\mu \nu}^e$
appearing in Eq.\ \rf{wcelec}. 
Note that the component $c_{00}^e$ 
cannot be bounded by such measurements,
since it remains unchanged as the
orientation of the magnetic field changes.
Together with the trace condition 
$c_{\pt{e}\mu}^{e\pt{\mu}\mu} = 0$,
this implies that a bound on the combination  
$|c_{11}^e + c_{22}^e|$
can also constrain $|c_{33}^e|$.

For possible diurnal variations 
of the electron cyclotron frequency,
an appropriate model-independent theoretical figure of merit
can be introduced as follows.
Define for the electron
\beq
\De^{e^-}_{\om_c} \equiv 
\fr {|{\cE}_{1,-1}^{e^-} - {\cE}_{0,-1}^{e^-}|}
{{\cE}_{0,-1}^{e^-}} 
\quad .
\label{Delediurnalom}
\eeq
An analogous definition 
could be introduced for the positron case.
Diurnal variations due to Lorentz violations
would appear as periodic fluctuations in
$\De^{e^-}_{\om_c}$.
We take their amplitude as a suitable figure of merit, 
$r^e_{\om_c, \rm diurnal}$.
In the context of the present framework,
we find
\beq
r^e_{\om_c, \rm diurnal} \approx 
{|c_{11}^e + c_{22}^e| \om_c}/{m_e}
\quad ,
\label{dcliman}
\eeq
again in the comoving Earth frame.
This figure of merit depends on the magnetic field
through $\om_c$,
which is appropriate because 
the associated types of level shift 
are explicitly dependent on $\om_c$,
as can be seen from Eq.\ \rf{elecdelE}.
As the applied field is increased,
the level shifts grow.

The results of Ref.\ \cite{schwin81}
can be used to estimate an upper bound
on $r^e_{\om_c, \rm diurnal}$.
During the 10-hour period in which data were taken,
the cyclotron frequencies varied 
by approximately 5 parts in $10^7$.
Attributing the whole of this to 
a hypothetical diurnal variation in $\om_c^{e^-}$ 
arising from the contribution $|c_{11}^e + c_{22}^e| \om_c$
produces an estimated upper bound 
\beq
r^e_{\om_c, \rm diurnal} \lsim 10^{-16}
\quad .
\label{cc}
\eeq
More recent techniques for stabilizing the magnetic field
might sharpen this bound by two orders of magnitude.
The bound could also be improved 
by monitoring the cyclotron frequencies over a longer time scale,
together with a search for signals with a diurnally related period.

\vglue 0.6cm
{\bf\noindent IV. PROTONS AND ANTIPROTONS}
\vglue 0.4cm

In this section,
we investigate some tests of CPT and Lorentz symmetry
using comparative Penning-trap
experiments with protons and antiprotons.
The discussion is divided into four subsections.
The first treats some issues for the underlying theory,
while the second and third consider experiments on
anomalous magnetic moments and charge-to-mass ratios,
respectively.
The fourth subsection examines comparative experiments
with hydrogen ions and antiprotons.

\vglue 0.6cm
{\bf\noindent A. Theory}
\vglue 0.4cm

At the level of the
SU(3)$\times$SU(2)$\times$U(1) standard model,
protons and antiprotons are composite particles
formed as bound states of quarks and antiquarks,
respectively.
Possible CPT- and Lorentz-violating effects
in the extension of the model
appear as perturbations involving the basic fields 
\cite{ck}.
For example,
a distinct set of parameters
$a_\mu$, $b_\mu$, $H_{\mu \nu}$, 
$c_{\mu \nu}$, $d_{\mu \nu}$
is assigned to each quark flavor,
and suitable combinations of these determine
the CPT- and Lorentz-violating features of the proton. 

For our present investigation
involving electromagnetic interactions
of protons and antiprotons in a Penning trap,
it suffices to work instead within
the usual effective theory 
in which the protons and antiprotons are regarded
as basic objects described 
by a four-component Dirac quantum field 
with dynamics governed by a minimally coupled lagrangian.
We therefore introduce effective parameters
$a_\mu^p$, $b_\mu^p$, $H_{\mu \nu}^p$, 
$c_{\mu \nu}^p$, $d_{\mu \nu}^p$
controlling possible CPT- and Lorentz-breaking
effects for the proton,
and we take the lagrangian to be the standard one
for proton-antiproton quantum electrodynamics
but extended to include 
possible small CPT- and Lorentz-violating terms. 
The corresponding Dirac equation has the form of Eq.\ \rf{dirac}.
The analysis of this model is analogous 
to the treatment presented in Sec.\ II.

We identify the Dirac hamiltonian $\hat H^p$ 
for the proton with $\hat H^q$ given in Eq.\ \rf{modDirac},
with perturbative terms as in Eq.\ \rf{Hint} 
except for superscripts $p$ 
on all CPT- and Lorentz-violating parameters
and the replacement $m \to m_p$
for the proton mass.
Similarly,
for the antiproton we identify
$\hat H^{\bar p} \equiv \hat H^{\bar q}$.
The wave functions for perturbative calculations
are well approximated as relativistic Landau eigenfunctions 
for protons and antiprotons.
We denote the associated energies,
including anomaly terms and other quantum effects
but excluding CPT- and Lorentz-breaking shifts,
by $E_{n,s}^p$ and $E_{n,s}^{\bar p}$.
The corresponding proton cyclotron and anomaly frequencies 
are defined as
\beq
\om_c = E_{1,+1}^{p} - E_{0,+1}^{p}
\quad ,
\quad\quad
\om_a = E_{0,-1}^{p} - E_{1,+1}^{p}
\quad .
\label{omdefsp}
\eeq
The CPT theorem implies they have the same values
as those of the antiproton.

Proceeding as in Sec.\ IIIA,
we can calculate perturbative energy corrections
that are first-order in CPT- and Lorentz-breaking parameters.
Contributions proportional to the magnetic field
are now suppressed by a factor of order $10^{-16}$.
Terms involving the axial or magnetron motions
are treated as before. 
Keeping only leading-order perturbations,
we find the corrections to the proton energies are
\bea
\de E_{n,\pm 1}^{p} &\approx& a_0^p \mp b_3^p
- c_{00}^p m_p \pm d_{30}^p m_p \pm H_{12}^p
\nonumber \\ &&
- \half (c_{00}^p + c_{11}^p +c_{22}^p) (2n + 1 \mp 1) \om_c 
\nonumber \\ &&
- \left( \half c_{00}^p + c_{33}^p \mp d_{03}^p \mp d_{30}^p
\right) \fr{p_z^2}{m_p} 
\quad .
\label{prodelE}
\eea
The energy shifts $\de E_{n,\pm 1}^{\bar p}$
for the antiproton can be obtained by the substitutions
$a_\mu^p \rightarrow - a_\mu^p$,
$d_{\mu \nu}^p \rightarrow - d_{\mu \nu}^p$,
$H_{\mu \nu}^p \rightarrow - H_{\mu \nu}^p$,
$E_{n,\pm 1}^{p} \rightarrow E_{n,\pm 1}^{\bar p}$,
and $(2n + 1 \mp 1) \rightarrow (2n + 1 \pm 1)$.
These results produce corrected cyclotron and anomaly frequencies.
At leading order in the CPT- and Lorentz-breaking quantities,
in the electromagnetic fields,
and in the fine-structure constant,
the modified frequencies are given by
\beq
\om_c^{p} = \om_c^{\bar p} \approx 
(1 - c_{00}^p - c_{11}^p - c_{22}^p) \om_c
\quad ,
\label{wcp}
\eeq
\beq
\om_a^{p} \approx \om_a
+ 2 b_3^p - 2 d_{30}^p m_p - 2 H_{12}^p
\quad , \qquad 
\om_a^{\bar p} \approx \om_a
- 2 b_3^p - 2 d_{30}^p m_p - 2 H_{12}^p
\quad .
\label{wap}
\eeq
Here,
$\om_c$ and $\om_a$ are the unperturbed frequencies
of Eq.\ \rf{omdefsp}.
Note that much of the discussion associated with the 
theoretical derivation in Sec.\ IIIA applies here. 
Note also that the ratio of proton and electron cyclotron frequencies 
is about $10^{-3}$,
whereas the proton and electron anomaly frequencies 
are roughly comparable in magnitude 
because the corresponding $g-2$ values
differ by a factor of about $10^3$.

\vglue 0.6cm
{\bf\noindent B. Anomalous Magnetic Moments}
\vglue 0.4cm

Currently,
the best measurements of the antiproton magnetic moment 
are accurate to only about 3 parts in $10^3$
and are extracted from experiments with exotic atoms
\cite{kriessl}.
In principle,
precision measurements of the anomalous magnetic moments of
protons and antiprotons could be obtained in Penning traps,
in analogy with the electron-positron experiments
discussed in Sec.\ IIIB,
provided sufficient cooling 
to temperatures below 4 K can be achieved.

A comparison of the experimental ratios
$2 \om_a^p / \om_c^p$ and
$2 \om_a^{\bar p} / \om_c^{\bar p}$ 
would then provide a stringent test of CPT and Lorentz violation.
No such experiments have been performed to date,
although the possibility has received some attention
in the literature
\cite{hw,qg}.

Using the present theoretical framework,
we can investigate the sensitivity 
of possible future $g-2$ experiments
to CPT and Lorentz violations.
To leading order,
we find the proton-antiproton differences
for the cyclotron and anomaly frequencies are 
\beq
\De \om_c^p \equiv \om_c^{p} - \om_c^{\bar p} = 0
\quad , \qquad
\De \om_a^p  \equiv \om_a^{p} - \om_a^{\bar p} = 4 b_3^p
\quad .
\label{delwap}
\eeq
Just as in the electron-positron case,
the leading-order signal for CPT breaking 
is thus an anomaly-frequency difference.
The corresponding figure of merit providing a well-defined
measure of CPT violation is
\beq
r^p_{\om_a}
\equiv \fr{|{\cal E}_{n,s}^{p} - {\cal E}_{n,-s}^{\bar p}|} 
{{\cal E}_{n,s}^{p}}
\quad ,
\label{rp}
\eeq
where the weak-field, zero-momentum limit is understood.
Within the present theoretical framework,
we find
\beq
r^p_{\om_a}
\approx |\De \om_a^p| / 2 m_p \approx |2 b_3^p|/m_p
\quad .
\label{rp2}
\eeq

Assuming an experiment could be made sensitive enough to measure
$\om_a^p$ and $\om_a^{\bar p}$
with a precision similar to that of electron $g-2$ experiments,
we can estimate the bound 
on $r^p_{\om_a}$ that would be attainable. 
For example,
supposing in analogy with the electron-positron experiments
that a frequency accuracy 
of about 2 Hz can be attained 
in the measurements of $\om_a^p$, $\om_a^{\bar p}$
and equality of $\om_c^p$, $\om_c^{\bar p}$ 
is observed to one part in $10^8$,
a bound of $|b_3^p| \lsim 10^{-15}$ eV
becomes possible.
This corresponds to an estimated bound on the figure of merit of
\beq
r^p_{\om_a}
\lsim 10^{-23}
\quad .
\label{rp3}
\eeq
It is evident that this experiment 
has the potential to provide  
a particularly stringent CPT bound in a baryon system.

Just as for the electron-positron case in Sec.\ IIIB,
experiments of this type could also bound 
diurnal variations in the average anomaly frequency.
An appropriate theoretical figure of merit in this case
can be introduced in terms of the quantity
\beq
\De^p_{\om_a} \equiv 
\fr {|{\cE}_{0,-1}^{p} - {\cE}_{1,+1}^{p}|}
{2{\cE}_{0,+1}^{p}} 
+
\fr {|{\cE}_{0,+1}^{\bar p} - {\cE}_{1,-1}^{\bar p}|}
{2{\cE}_{0,-1}^{\bar p}} 
\quad .
\label{Delpdiurnal}
\eeq
The figure of merit is the amplitude 
$r^p_{\om_a, \rm diurnal}$
of diurnal variations
observed in $\De^p_{\om_a}$.
In the present framework,
these depend on Lorentz-violating but CPT-preserving terms,
and we find 
\beq
r^p_{\om_a, \rm diurnal}
\approx | d_{30}^p m_p + H_{12}^p|/m_p
\quad ,
\label{omsump}
\eeq
in the comoving Earth frame.
Assuming observations confine diurnal variations of  
the anomaly-frequency average to within a 1 kHz band as before,
we obtain an estimated bound on the figure of merit of 
\beq
r^p_{\om_a, \rm diurnal}
\lsim 10^{-21}
\quad .
\label{dHp}
\eeq

\vglue 0.6cm
{\bf\noindent C. Charge-to-Mass Ratios}
\vglue 0.4cm

Experiments confining single protons and antiprotons
in an open-access Penning trap provide high-precision
comparisons of their cyclotron frequencies
\cite{gg1},
yielding the limit 
$|\De\om_c^p| / \om_c^p \lsim 10^{-9}$.
The corresponding conventional figure of merit
$r_{q/m}^p$
and its current bound 
are given in Eq.\ \rf{rqmp}.

Within the present theoretical framework,
Eq.\ \rf{prodelE} demonstrates that 
the CPT- and Lorentz-violating terms introduce 
nonzero energy-level shifts,
even in the weak-field zero-momentum limit.
The perturbations of the cyclotron frequencies
are given in Eq.\ \rf{delwap}.
To leading order,
the proton and antiproton cyclotron frequencies 
are independent of CPT-violating quantities,
just as for the electron-positron case
discussed in Sec.\ IIIC.
As the cyclotron frequencies are unaffected 
even if CPT is broken,
a comparison of these frequencies 
would represent a misleading measure of CPT violation.
For example,
the figure of merit $r_{q/m}^p$ in Eq.\ \rf{rqmp},
which is proportional to
the frequency difference $\De \om_c^p$,
may vanish even though the model contains explicit CPT violation.

The Lorentz-breaking but CPT-preserving parameters
induce identical shifts 
in the proton and antiproton cyclotron frequencies.
In analogy with the electron-positron case,
this indicates that the frequency difference $\De \om_c^p$
would be an inappropriate measure of Lorentz violation.

Another possibility is the occurrence of diurnal variations
in the cyclotron frequencies,
which could be induced by the Earth's rotation 
during the course of an experiment.
Such variations would arise in the present context
from the dependence of the cyclotron frequencies on
the components $|c_{11}^p + c_{22}^p|$ of $c_{\mu \nu}^p$.
As discussed for the electron-positron case in Sec.\ IIIC,
the unobservability of the component $c_{00}^p$ 
means that a bound on $|c_{11}^p + c_{22}^p|$ 
can also constrain $|c_{33}^p|$.

A suitable theoretical figure of merit 
can be introduced in analogy with the electron-positron case.
Define for the proton 
\beq
\De^{p}_{\om_c} \equiv 
\fr {|{\cE}_{1,-1}^{p} - {\cE}_{0,-1}^{p}|}
{{\cE}_{0,-1}^{p}} 
\quad .
\label{Delpdiurnalom}
\eeq
The figure of merit is the amplitude 
$r^p_{\om_c, \rm diurnal}$
of periodic fluctuations in $\De^{p}_{\om_c}$.
In the comoving Earth frame,
we find
\beq
r^p_{\om_c, \rm diurnal} 
\approx {|c_{11}^p + c_{22}^p| \om_c}/{m_p} 
\quad .
\label{dclimanp}
\eeq
As for the corresponding electron-positron case,
the appearance of $\om_c$ implies 
that the value of this figure of merit 
depends on the magnetic field.
This is appropriate,
since the associated level shifts
in Eq.\ \rf{prodelE}
also explicitly depend on $\om_c$.

A crude estimated upper bound on 
$r^p_{\om_c, \rm diurnal}$
can be obtained from the data in Ref.\ \cite{gg1},
which represent alternate measurements 
of proton and antiproton cyclotron frequencies 
over a 12-hour period.
The slow drifts in these frequencies 
are confined to a band of approximate width 2 Hz.
This suggests an upper bound on a possible diurnal variation
in $r^p_{\om_c, \rm diurnal}$
arising from the contribution proportional to 
$|c_{11}^p + c_{22}^p|$,
given by
\beq
r^p_{\om_c, \rm diurnal} \lsim 10^{-24}
\quad .
\label{ccp}
\eeq
Note that diurnal fluctuations in
the antiproton cyclotron frequency could be treated similarly.

The bound \rf{ccp} is better than 
the corresponding one for electrons
and positrons given in Eq.\ \rf{cc}.
It might be sharpened through detailed analysis 
of the experimental data,
perhaps including a fit for diurnal variations
and compensation for known correlations 
with temperature fluctuations in the experimental hall.

\vglue 0.6cm
{\bf\noindent D. Experiments with Hydrogen Ions}
\vglue 0.4cm

When protons and antiprotons are interchanged 
in the Penning-trap experiments of Ref.\ \cite{gg1},
the associated reversal of the electric field
can lead to offset potentials affecting differently 
the proton and antiproton cyclotron frequencies.
In an ingenious recent experiment
\cite{gg},
Gabrielse and coworkers have addressed this issue 
by comparing antiproton cyclotron frequencies
with those of an $H^-$ ion instead of a proton.
The equality of the charges means the same trap
and fields can be used,
and the experiment also allows relatively rapid interchanges 
between hydrogen ions and antiprotons.
The expected theoretical value of the difference
$\De \om_c^{H^-}
\equiv\om_c^{H^-} - \om_c^{\bar p}$
can be obtained in the context of conventional quantum theory 
using established precision measurements of the electron mass
and the $H^-$ binding energy.
Comparison of this theoretical value
with the experimental result for 
$\De \om_c^{H^-}$
is expected to provide a symmetry test
with a precision of about one part in $10^{10}$.

Understanding the implications of this experiment 
within the present theoretical framework
requires a description of the electromagnetic
interactions of the hydrogen ion in a Penning trap 
in the presence of possible CPT and Lorentz violation.
A hydrogen ion can be regarded as a charged composite fermion, 
so its electromagnetic interactions can be discussed 
within an effective spinor electrodynamics
producing a Dirac equation of the form \rf{dirac}
for a fermion of mass $m_{H^-}$.
The corresponding effective CPT- and Lorentz-breaking parameters
are denoted
$a_\mu^{H^-}$, $b_\mu^{H^-}$, $H_{\mu \nu}^{H^-}$, 
$c_{\mu \nu}^{H^-}$, $d_{\mu \nu}^{H^-}$.
The theoretical treatment then proceeds as in Sec.\ II.

For a hydrogen ion in a Penning trap,
we obtain the leading-order energy shifts 
from CPT and Lorentz breaking
following the method in Secs.\ IIIA and IVA.
We find
\bea
\de E_{n,\pm 1}^{H^-} &\approx& a_0^{H^-} \mp b_3^{H^-}
- c_{00}^{H^-} m_{H^-} 
\pm d_{30}^{H^-} m_{H^-} \pm H_{12}^{H^-}
\nonumber \\ &&
- \half (c_{00}^{H^-} + c_{11}^{H^-} +c_{22}^{H^-})
(2n + 1 \pm 1) \om_c^{H^-} 
\nonumber \\ &&
- \left ( c_{00}^{H^-} - c_{33}^{H^-} 
\mp d_{03}^{H^-} \mp d_{30}^{H^-}\right) 
\fr{p_z^2}{m_{H^-}} 
\quad .
\label{HdelE}
\eea
The $H^-$ cyclotron frequency is therefore shifted 
from its value $\om_{c}^{H^-}$ 
in the absence of Lorentz violation
to a perturbed value $\om_{c,\rm pert}^{H^-}$ given by
\beq
\om_{c,\rm pert}^{H^-} 
\approx 
(1 - c_{00}^{H^-} - c_{11}^{H^-} - c_{22}^{H^-}) \om_{c}^{H^-} 
\quad .
\label{delomH}
\eeq
Much of the discussion in Secs.\ IIIA and IVA 
concerning the corresponding theoretical derivations
also applies here.

The above result can be used to obtain limits 
on Lorentz-violating quantities for hydrogen ions and protons.
Denote as before the difference between the cyclotron frequencies
of the hydrogen ion and the antiproton by 
$\De \om_c^{H^-}$.
Then,
the component $\De \om_{c,\rm th}^{H^-}$ 
of $\De \om_c^{H^-}$
that is determined theoretically to arise 
purely from CPT- and Lorentz-violating effects
can be obtained from Eqs.\ \rf{wcp} and \rf{delomH}.
We find
\beq
\De \om_{c,\rm th}^{H^-}
\approx
( c_{00}^p + c_{11}^p + c_{22}^p) \om_c
-(c_{00}^{H^-} + c_{11}^{H^-} + c_{22}^{H^-}) \om_{c}^{H^-} 
\quad .
\label{delwthbasic}
\eeq
As before, $\om_c$ is the proton-antiproton
cyclotron frequency in the absence of CPT or Lorentz perturbations.

The definition of a model-independent figure of merit
proceeds in analogy with the treatments in preceding sections.
We introduce the quantity
\beq
\De^{H^-}_{\om_c} \equiv 
\fr {|{\cE}_{1,-1}^{H^-} - {\cE}_{0,-1}^{H^-}|}
{2{\cE}_{0,-1}^{H^-}} 
- \fr {|{\cE}_{1,-1}^{\bar p} - {\cE}_{0,-1}^{\bar p}|}
{2{\cE}_{0,-1}^{\bar p}} 
\quad .
\label{DelHomc}
\eeq
As defined,
$\De^{H^-}_{\om_c}$ is nonzero 
even if CPT and Lorentz symmetry is preserved.
To obtain a measure that vanishes in the exact symmetry limit,
we remove from the hydrogen-ion terms in $\De^{H^-}_{\om_c}$
the conventional contributions arising from the differences
between the $H^-$ ion and a proton:
the masses of the two electrons and the binding energy.
The result is a suitable figure of merit for Lorentz violation,
denoted by $r^{H^-}_{\om_c}$.
The calculations leading to Eq.\ \rf{delwthbasic}
imply that within the present framework 
\beq
r^{H^-}_{\om_c} \approx
|\De \om_{c,\rm th}^{H^-}| / {m_p} 
\quad .
\label{rHomc}
\eeq

It is plausible that a precision of about
one part in $10^{10}$ could be attained 
in measurements of the ratio 
$|\De \om_c^{H^-}|/\om_c^{H^-}$.
Suppose the observed value agrees with conventional
theory to within a certain accuracy.
Then,
this accuracy must be larger 
than the predicted shift ratio 
$|\De \om_{c,\rm th}^{H^-}|/\om_c^{H^-}$.
We thus obtain an estimated bound of 
\beq
r^{H^-}_{\om_c} \lsim 10^{-25}
\quad 
\label{ccHp}
\eeq
that might be attained in this class of experiment.

The above results involve a combination of the
Lorentz-violating quantities for hydrogen ions and protons.
However,
all the effective CPT- and Lorentz-breaking parameters
for a hydrogen ion are determined 
by appropriate combinations of the corresponding parameters 
for its constituents.
Lowest-order perturbation theory can be used to find
approximations to these relationships.
The wave function of the hydrogen ion
can be treated as a product of a proton wave function 
and a two-electron wave function,
and the corresponding 
net CPT- and Lorentz-breaking energy shifts 
induced for the hydrogen ion can be estimated,
neglecting nonperturbative issues involving binding effects.

In this approximation,
we find 
\beq
c_{\mu\mu}^{H^-} \approx 
c_{\mu\mu}^p 
+ (c_{\mu\mu}^e - c_{\mu\mu}^p ) \fr{2m_e}{m_p} 
\quad ,
\label{hrel}
\eeq
where no sum is implied on repeated indices.
Substitution of this result into Eq.\ \rf{delwthbasic} gives
\beq
\De \om_{c,\rm th}^{H^-}
\approx
(c_{00}^{p} + c_{11}^{p} +c_{22}^{p}) 
( \om_c - \om_c^{H^-} )
- \fr{2m_e}{m_p} 
(c_{00}^{e} + c_{11}^{e} + c_{22}^{e}
-c_{00}^{p} - c_{11}^{p} -c_{22}^{p}) 
\om_{c}^{H^-} 
\quad .
\label{delwth}
\eeq
This result implies
that the bound in Eq.\ \rf{ccHp}
constrains a combination of 
Lorentz-violating but CPT-preserving quantities,
including $c_{00}^e$ and $c_{00}^p$.
The latter would be inaccessible through the other experiments 
considered in the present work. 
Moreover,
this experiment does not require searching 
for diurnal variations in the cyclotron frequency,
which means potential systematics
associated with diurnal field drifts are eliminated.

We remark in passing that in principle
anomaly-frequency comparisons
of $H^-$ and antiprotons could also be envisaged.
Leaving aside experimental issues,
the theoretical motivation for such experiments
seems somewhat lacking.
One point is that perturbative calculation indicates
$b^{H^-}_\mu \approx b^p_\mu$,
so bounds that might be obtained in this way
would also be accessible in 
the experiments mentioned in Sec.\ IIIB.

\vglue 0.6cm
{\bf\noindent V. SUMMARY}
\vglue 0.4cm

In this paper,
we have used a general theoretical framework
based on an extension of the standard model
and quantum electrodynamics
to establish and investigate possible signals
of CPT and Lorentz breaking
in certain Penning-trap experiments.
We have focused on leading-order limits  
arising from high-precision measurements 
of anomaly and cyclotron frequencies.
Table I summarizes our results.

Our estimated bounds from experiments 
with the electron-positron system
are given in Eqs.\ \rf{relim}, \rf{dHlim}, and \rf{cc}.
Bounds from the proton-antiproton system are in
Eqs.\ \rf{rp3}, \rf{dHp}, and \rf{ccp},
while a bound from the $H^-$-antiproton system
is given in Eq.\ \rf{ccHp}.

Sharp tests of CPT symmetry emerge from $g - 2$ experiments.
We have introduced appropriate figures of merit
with attainable bounds of approximately $10^{-20}$ 
using current methods in the electron-positron case
and of $10^{-23}$ for a plausible experiment 
with protons and antiprotons.
Other experimental signals 
originating from CPT-preserving Lorentz violations could occur,
involving possible diurnal variations in frequency measurements.
These could produce bounds at the level of 
$10^{-18}$ in the electron-positron system
and $10^{-21}$ in the proton-antiproton system.

In contrast,
comparative measurements of cyclotron frequencies
for particles and antiparticles are insensitive 
to leading-order effects from 
CPT breaking within the present framework.
However,
diurnal variations of cyclotron frequencies
and comparative measurements of cyclotron frequencies
for hydrogen ions and antiprotons are affected by different 
CPT-preserving Lorentz-violating quantities.
These experiments could generate bounds 
on various dimensionless figures of merit 
at the level of
$10^{-16}$ in the electron-positron system, 
$10^{-24}$ in the proton-antiproton system,
and $10^{-25}$ using the $H^-$-antiproton system.

\vglue 0.6cm
{\bf\noindent ACKNOWLEDGMENTS}
\vglue 0.4cm

This work is supported in part 
by the Department of Energy
under grant number DE-FG02-91ER40661 
and by the National Science Foundation
under grant number PHY-9503756.

\vglue 0.6cm

\newpage
\bigskip
\begin{tabular}{|l|l|c|c|l|c|} 
\hline\hline
\multicolumn{2}{|c|}{Experiment}
  & Fig.\ of Merit & Est.\ Bound & Parameters & Eq.\
\\
\hline\hline
$e^- e^+$ & $\om_a$ comparison & $r^e_{\om_a}$ & $10^{-20}$ & 
   $b_j^e$ & \rf{relim} \\[2mm]
\cline{2-6}
 & diurnal $\om_a$ variation & $r^e_{\om_a, \rm diurnal}$
 & $10^{-18}$ & $d_{j0}^e,\  H_{jk}^e$ &  \rf{dHlim} \\[2mm]
\cline{2-6}
 & diurnal $\om_c$ variation & $r^e_{\om_c, \rm diurnal}$
 & $10^{-16}$ & $c_{jj}^e$ &  \rf{cc}\\[2mm]
\hline
$p \bar p$ & $\om_a$ comparison & $r^p_{\om_a}$
 & $10^{-23}$ & $b_j^p$ & \rf{rp3} \\[2mm]
\cline{2-6}
 & diurnal $\om_a$ variation & $r^p_{\om_a, \rm diurnal}$
 & $10^{-21}$ &  $d_{j0}^p, \ H_{jk}^p$ & \rf{dHp} \\[2mm]
\cline{2-6}
 & diurnal $\om_c$ variation & $r^p_{\om_c, \rm diurnal}$
 & $10^{-24}$ & $c_{jj}^p$ &  \rf{ccp}\\[2mm]
\hline
$H^- \bar p$& $\om_c$ comparison & $r^{H^-}_{\om_c}$
 & $10^{-25}$ & $c_{\mu\mu}^{H^-}$, $c_{\mu\mu}^{p}$ & 
   \rf{ccHp} \\[2mm]
\hline
\hline
\end{tabular}
\normalsize

\bigskip
{\bf Table 1.} 
Estimated CPT- and Lorentz-violating bounds 
for electron-postron, proton-antiproton, 
and $H^-$-antiproton experiments.
The first two columns specify the type of experiment.
The third column lists figures of merit,
while the fourth gives the corresponding bounds
estimated from current or future experiments.
The fifth column shows which of the quantities in Eq.\ \rf{dirac}
enter the constraint.
Entries in the final column are the numbers for the equations 
in the text where the bound is presented.

\bigskip

\end{document}